# A model combining spectrum standardization and dominant factor based partial least square method for carbon analysis in coal by laser-induced breakdown spectroscopy

Xiongwei Li,[1] Zhe Wang[*,1] Yangting Fu,[1] Zheng Li,[1] Weidou Ni[1]

[1]*State Key Lab of Power Systems, Department of Thermal Engineering, Tsinghua-BP Clean Energy Center, Tsinghua University, Beijing, China*

**Abstract:** Successful quantitative measurement of carbon content in coal using laser-induced breakdown spectroscopy (LIBS) is suffered from relatively low precision and accuracy. In the present work, the spectrum standardization method was combined with the dominant factor based partial least square (PLS) method to improve the measurement accuracy of carbon content in coal by LIBS. The combination model employed the spectrum standardization method to convert the carbon line intensity into standard state for more accurately calculating the dominant carbon concentration, and then applied PLS with full spectrum information to correct the residual errors. The combination model was applied to the measurement of carbon content for 24 bituminous coal samples. The results demonstrated that the combination model could further improve the measurement accuracy compared with both our previously established spectrum standardization model and dominant factor based PLS model using spectral area normalized intensity for the dominant factor model. For example, the coefficient of determination ($R^2$), the root-mean-square error of prediction (RMSEP), and the average relative error (ARE) for the combination model were 0.99, 1.75%, and 2.39%, respectively; while those values for the spectrum standardization method were 0.83, 2.71%, and 3.40%, respectively; and those values for the dominant factor based PLS model were 0.99, 2.66%, and 3.64%, respectively.

## 1  Introduction

Carbon content is one of the most important indexes that reflect coal quality [1]. On-line measurement of carbon content in coal contributes to the quick estimation of coal's calorific value, and therefore is very useful for power plant to realize combustion optimization and coal pricing in real time [2-4]. The laser-induced breakdown spectroscopy (LIBS) is a very promising technology for on-line coal analysis, for its advantages include rapid and in situ analysis, no or minimal sample preparation, simultaneous multi-element measurement, and so on [5-7]. Although a number of studies on coal analysis by LIBS have been performed [8-12], there are very few reported researches about the quantitative determination of non-metallic elements (e.g., C, H, O, and N) in coal [10-12]. The measurement precision and accuracy of these elements are not satisfying due to multiple factors, such as matrix effects, variations in experimental condition, and the complex physical and chemical processes of the laser-induced plasma from its generation to its expansion into the ambient gas [13-16]. Some experimental modification methods, such as dual-pulse [17] and fast-discharge [18, 19], have been proposed to improve the analytical performance of LIBS. Another effective method without increasing the experimental complexity is the data processing method. Various mathematical and statistical approaches have been applied to process the spectral data. The commonly used method is to normalize the signal by background emission or spectral area [20-22], which, however, cannot effectively improve the measurement precision and accuracy of carbon content in coal [23].

In our previous work, a modified spectrum standardization model was proposed to achieve more reproducible and accurate results for the quantitative measurement carbon content in coal [24]. The modified model utilized the emission intensity of $C_2$ to compensate for the diminution of atomic carbon emission in high volatile coal samples caused by matrix effect, and then converted the compensated carbon line intensities into values at a standard state to further compensate for the line intensities' fluctuations caused by the variations of plasma properties, including plasma temperature, electron density, and total number density of carbon. The application of the modified model for the measurement of carbon concentrations in 24 bituminous coal samples showed that the model could obviously improve both measurement precision and accuracy.

[*] Corresponding author. Tel: +86 10 62795739; fax: +86 10 62795736
   Email address: zhewang@tsinghua.edu.cn

We also proposed in our previous study a dominant factor based PLS model to determine carbon concentrations in bituminous coals [23]. The dominant factor based PLS model utilized the most related spectral intensities to establish the dominant factor for calculating the major portion of carbon concentration, then further compensated for the remaining residual errors using PLS with entire spectrum information. The dominant factor endowed the proposed model with physical background as the traditional uni-variate model possessed, and therefore the proposed model was able to provide more accurate results than the conventional PLS by reducing the interference of unrelated noise to some extent.

Till now, the two different methods have always been applied for LIBS quantitative analysis individually. In fact, these two methods can be naturally combined together: the spectrum standardization method can be regarded as a data pretreat method, which is able to provide a more accurate dominant factor model for the dominant factor based PLS model. In the present work, the two models were combined together for the measurement of carbon concentration in coal.

## 2 Method introduction

In the previously proposed dominant factor based PLS model [25], the major part of elemental concentration was explained by explicitly extracting the most related spectral information such as the characteristic line intensities of the measured element. The explicitly extracted expression is called "the dominant factor" since it takes the dominant part of final calculated concentration values. After extracting the dominant factor, there is still a deviation between the real elemental concentration and the value calculated by the dominant factor, due to the fluctuations of plasma temperature, electron number density, inter-element interference, or other unknown deviation factors. As the entire spectrum contains some useful information about the deviation sources, it was utilized to further minimize the deviation by the PLS algorithm.

In our previous work [23, 25], it was found that the final model results were largely dependent on the dominant factor. A more accurate dominant factor contributed to the improvement of the model. Our previously established spectrum standardization model compensated for the diminution of atomic carbon emission in high volatile coal samples as well as the line intensities' fluctuations caused by the variations of plasma properties [24]. Therefore, a more accurate dominant factor model can be just provided by the spectrum standardization model, which is basically a uni-variate model with physical background. That is, the standardized carbon line intensities can be naturally utilized to establish a more accurate dominant factor for the dominant factor based PLS model to improve the quantitative analytical performance. The procedure to establish the dominant factor based PLS model is briefly described as follows, and thereafter the procedure to establish the spectrum standardization based PLS method is introduced.

The first step to establish the dominant factor based PLS model is to extract the main relationship between the elemental concentration and the characteristic intensity of the measured element. It is,

$$C_i = f(I_i) \qquad (1)$$

where $C_i$ is the elemental concentration of the measured element $i$, $I_i$ is the characteristic intensity of the measured element $I$, $f(I_i)$ could be non-linear if the self-absorption effect cannot be neglected.

As inter-element interference may be a major source for the deviation between the real element concentration and the value calculated with Eq.1, the second step is to model the inter-element interference to minimize the deviation by best curve-fitting technology with nonlinear equation. That is,

$$C_i^{'} = f(I_i) + g(I_j) \tag{2}$$

where $C_i^{'}$ is the elemental concentration calculated from the dominant factor that considers self-absorption and inter-element interference, $I_j$ is the characteristic line intensity of the influencing element $j$, $g(I_j)$ is the function to describe inter-element interference.

After dominant factor extraction, the PLS method is applied to further compensate for the deviation using the entire spectrum information. Then the final expression of the model is

$$C_i^{''} = f(I_i) + g(I_j) + b_0 + b_1 x_1 + ... + b_n x_n \tag{3}$$

where $C_i^{''}$ is the calculated elemental concentration of the dominant factor based PLS model, $x_1$, $x_2$, ..., $x_n$ are the spectral intensities at different wavelengths, $b_1$, $b_2$, ..., $b_n$ are the regression coefficients. For more details of the dominant factor based PLS model, please referred to our previous work [25].

In the combination model, the dominant factor, which is major part of elemental concentration, is extracted by standardized carbon line intensity. The standardized carbon line intensity is calculated from our previous spectrum standardization method [24]. It is,

$$I_{ij}(n_{s0}, T_0, n_{e0}) = I_{ij} + \sum_{i=1}^{k} b_{1i} I_{Ti} C + b_2 C + b_3 \left( \ln\left(\frac{I_2}{I_1}\right) - \left(\ln\left(\frac{I_2}{I_1}\right)\right)_0 \right) C + b_4 (\Delta\lambda_{stark} - (\Delta\lambda_{stark})_0) C \tag{4}$$

where $C$ is the carbon concentration, $I_{ij}(n_{s0}, T_0, n_{e0})$ is the standard carbon line intensity, $I_{ij}$ is the compensated carbon line intensity obtained from the linear combination of the emission intensity of atomic carbon and the emission intensity of molecular carbon, $I_{Ti}$ is a segmental spectral area, $I_2/I_1$ is the intensity ratio of a pair of lines, $\Delta\lambda_{stark}$ is full width of half maximum (FWHM) of the $H_\alpha$ spectral line through Stark broadening, both $[\ln(I_2/I_1)]_0$ and $(\Delta\lambda_{stark})_0$ are calculated from all the measured spectra's average to indicate their standard state values, $b_{1i}$, $b_2$, $b_3$, and $b_4$ are constants calculated from an iterative regression process .

By extracting the main relationship between the standardized line intensity of carbon and the carbon concentration, the dominant factor model can be established as

$$C = k I_{ij}(n_{s0}, T_0, n_{e0}) + b \tag{5}$$

where $C$ is the carbon concentration, $I_{ij}(n_{s0}, T_0, n_{e0})$ is the standard carbon line intensity, $k$ and $b$ are constants calculated from the regression process.

To further correct the imperfectness of the dominant factor, inter-element interference and other unknown factors, the deviation between the real element concentration and the value calculated with Eq. 5 was compensated by the entire spectrum information with PLS method. The final expression of the combined model is

$$C_1 = k I_{ij}(n_{s0}, T_0, n_{e0}) + b_0 + b_1 x_1 + ... + b_n x_n \tag{6}$$

where $C_1$ is final calculated carbon concentration of the combination model, $x_1$, $x_2$, ..., $x_n$ are the spectral intensities at different wave lengths, and $b_0$, $b_1$, $b_2$, ..., $b_n$ are the regression coefficients calculated by the PLS algorithm.

## 3 Experimental setup

Spectrolaser 4000 system (XRF Scientific, Australia) was used for the experiments. The experimental arrangement has been described previously [23]. Briefly, a Q-switched Nd:YAG laser emitting at 532 nm with a pulse duration of 5 ns was used in the experiment, and the laser energy was adjusted to be 120 mJ/pulse. The laser beam was focused onto the sample surface and the spot diameter is 200 μm. Four Czerny-Turner spectrographs and charge coupled device (CCD) detectors covered an overall range (nm) from 190 to 310, 310 to 560, 560 to 770, and 770 to 950, respectively, with a nominal resolution of 0.09 nm. The gate delay time was adjusted to be 2 μs, and the integration time was fixed at 1 ms.

Twenty-four standard bituminous coal samples, which were certified by the China Coal Research Institute, were used in the experiment. Carbon concentrations in these coal samples ranged from 42% to 82% (Table 1). The powder of each coal sample were placed into a small aluminum pellet die ($\phi$=30 mm, $h$=3 mm) and then were pressed with the pressure of 20 tons. The samples were mounted on an auto-controlled *X-Y* translation stage and exposed to air. The samples were divided into the calibration and validation sets. To ensure a wide range and even concentration distribution in both sets, all samples were first arranged by their C concentrations, and then one of every three samples was chosen for validation.

**Table 1.** Carbon concentrations of 24 coal samples.

|                  | No.     | 1     | 2     | 3     | 4     | 5     | 6     | 7     | 8     |
|------------------|---------|-------|-------|-------|-------|-------|-------|-------|-------|
| **Calibration Set** | C (%)   | 47.12 | 52.61 | 53.77 | 54.72 | 58.12 | 59.84 | 67.18 | 67.77 |
|                  | No.     | 9     | 10    | 11    | 12    | 13    | 14    | 15    | 16    |
|                  | C (%)   | 70.45 | 74.7  | 76.69 | 77.28 | 78.64 | 79.02 | 79.98 | 81.54 |
| **Validation Set** | No.     | 17    | 18    | 19    | 20    | 21    | 22    | 23    | 24    |
|                  | C (%)   | 53.42 | 55.67 | 59.91 | 72.71 | 75.96 | 78.58 | 79.7  | 81.45 |

Twenty-five locations were measured for each pellet and each location was fired twice. The first shot of 150 mJ is to remove any contaminant, and the second shot of 120 mJ is used for analysis. The aerosol particles produced from each laser shot was blow off to eliminate aerosol influence on the signal. Background was subtracted from each spectrum to reduce the systematic signal fluctuation. The intensity was defined as the integration of channel readings of an emission line above the background. The system was warmed up for at least 1 h to ensure the thermal stability of the instruments.

## 4  Results and discussion

The performance of the combination model is evaluated by comparing with two baseline models. The spectrum standardization method was chosen as one of the baselines since it provided the dominant factor of the proposed model. The second baseline was our previous dominant factor based PLS model, which applied characteristic line intensities with the spectral area normalization for the dominant factor.

Three parameters are chosen to evaluate the performance of the proposed model. These parameters include the $R^2$ of calibration curve, the root mean square error of prediction (RMSEP) of mass concentration, and the average relative error (ARE) of predicted mass concentrations. The $R^2$ can assess the quality of the data points that are used to establish the calibration model. The RMSEP and ARE can indicate the accuracy of predictions by the models. In addition, the number of principle components in the PLS model was determined by the leave-one-out cross validation (LOO-CV) method to avoid noise over-fitting.

The integrated intensity of C(I) 247.856 nm was selected to establish the uni-variate calibration model. The measured C(I) 247 nm line intensity does not have a good linearity with the carbon concentration due to the strong matrix effect, so it was normalized with the segmental spectral area [23]. As shown in Fig.1, the $R^2$ value of the calibration plot for the segmental area normalization method is 0.75. The RMSEP and ARE from the segmental normalization method are 3.77% and 4.10%, respectively.

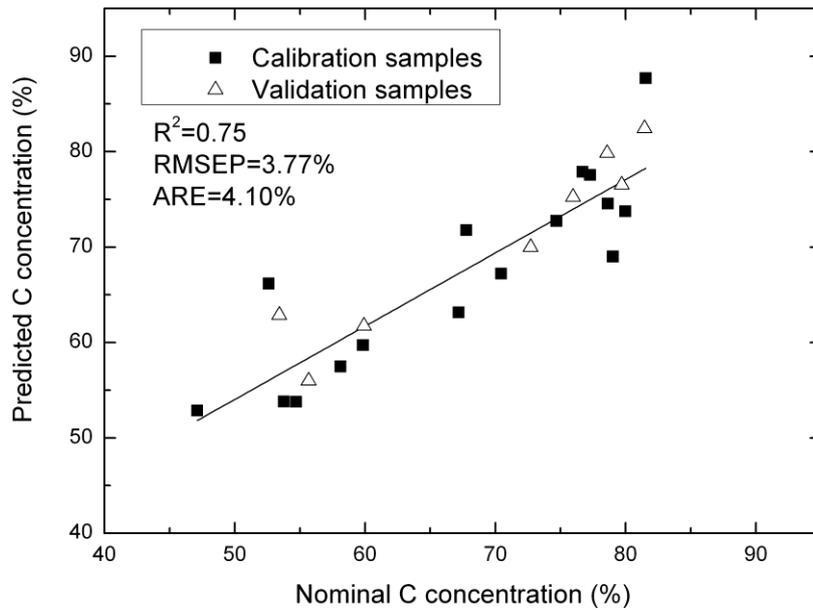
**Figure 1.** Calibration and validation results of segmental spectral area normalization method

Fig. 2 demonstrates the calibration and validation results of the spectrum standardization method. Compared with the segmental normalization method, the $R^2$ is increased to 0.83, RMSEP is lowered to 2.71%, and ARE is lowered to 3.40%. The improvement indicates that the matrix effect can be better corrected by the standardization method which compensates for the fluctuations of spectral line intensities resulted from the variation of plasma parameters ($T$, $n_e$, and $n_s$).

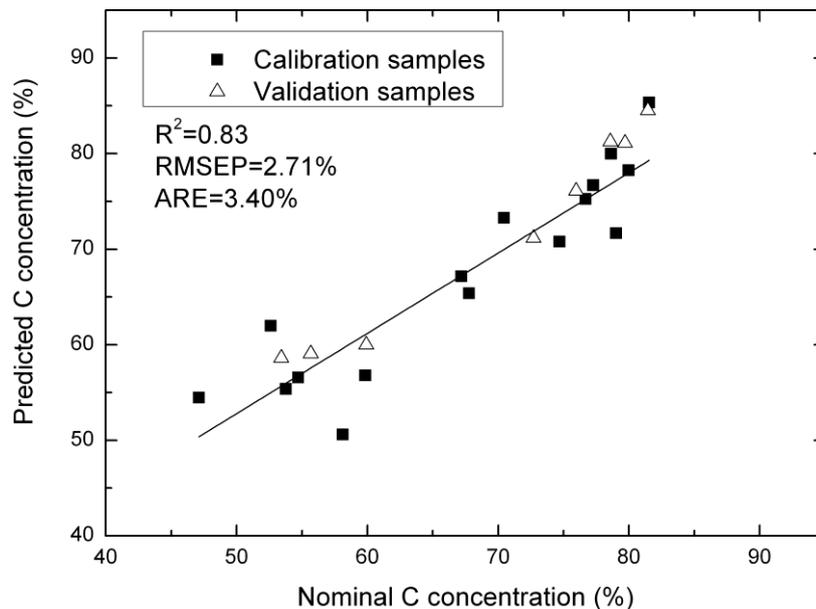
**Figure 2.** Calibration and validation results of the spectrum standardization method

Fig. 3 shows the calibration and validation results of the PLS model with the normalized carbon line intensity as the dominant factor. As shown in Fig.3, $R^2$ is 0.99, RMSEP is 2.66%, and ARE is 3.64%. Compared with the segmental spectral area normalization, the improvement in $R^2$ and the reduction in RMSEP and ARE indicate that the prediction accuracy can be improved by utilizing the full spectrum information to further compensate the deviations.

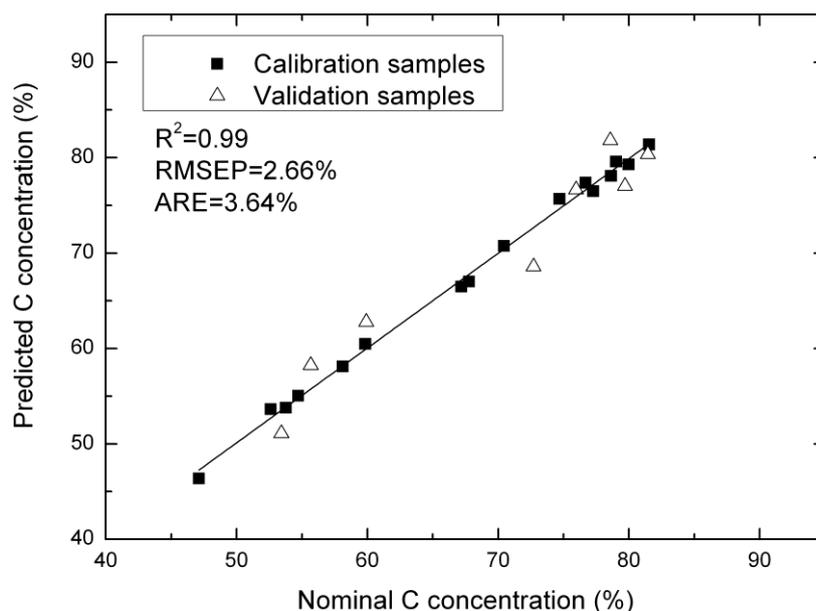

**Figure 3.** Calibration and validation results of the dominant factor based PLS model with the normalized spectral line intensity as the dominant factor

Figure 4 demonstrates the calibration and validation results of the combination model. As seen, $R^2$ is 0.99, RMSEP is 1.75%, and ARE is 2.39%. All of the three parameters from the combination model were found to be much better than those from the spectrum standardization method. This also suggested the introduction of the full spectrum information to compensate the residuals can effectively improve the predictive accuracy. Besides, the prediction of the present model was more accurate than that of the PLS model with the normalized spectral line intensity as the dominant factor judging from RMSEP and ARE. This confirmed the advantage of utilizing the standardized spectral intensity as the dominant factor in improving the predictive accuracy of model.

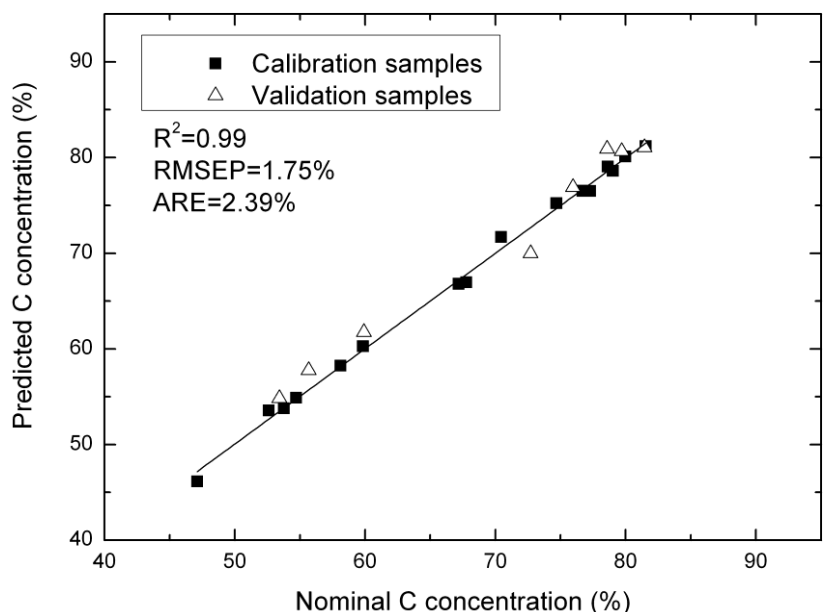

**Figure 4.** Calibration and validation results of the combination model

Table 2 lists the summary of analytical results of different models. As seen, the combination model performed the best among the four models judging from all the parameters, showing the advantage of the proposed model in improving the quantitative analysis of carbon content in coal by LIBS.

**Table 2.** Summary of performance of different models

| Models | $R^2$ | RMSEP/% | ARE/% |
|---|---|---|---|
| Segmental area normalization | 0.75 | 3.77 | 4.10 |
| Spectrum standardization | 0.83 | 2.71 | 3.40 |
| PLS model with the normalized spectral line intensity as the dominant factor | 0.99 | 2.66 | 3.64 |
| Combination model | 0.99 | 1.75 | 2.39 |

# 5 Conclusions

The previously established spectrum standardization method can obviously improve the measurement precision and accuracy by converting the measured carbon line intensity to the standard carbon line intensity. The proposed combination model employed the standard carbon line intensity as the dominant factor to explain the main carbon concentration, and further corrected the deviation by the full spectrum information with PLS algorithm. The assay of the carbon concentration of 24 bituminous coal samples by the proposed model showed an obvious improvement in measurement accuracy compared with spectrum standardization method as well as the dominant factor based PLS model with the spectral area normalization for the dominant factor model.

# Acknowledgement

The authors are grateful for the financial support from the National Natural Science Foundation of China (Grant No. 51276100) and National Basic Research Program (973 Program) (No. 2013CB228501).